\documentclass{aastex}
\usepackage{spr-astr-addons}
\usepackage{url}
\usepackage{color}

\usepackage{graphicx}
\begin{document}


\title{Electromagnetic Fields of Slowly Rotating Compact Magnetized Stars in Braneworld}

\author{V.S. Morozova\altaffilmark{1,2}} \and
\author{B.J. Ahmedov\altaffilmark{1,2,3}}

\altaffiltext{1}{Ulugh Begh Astronomical Institute,
Astronomicheskaya 33, Tashkent 100052, Uzbekistan}
\altaffiltext{2}{Institute of Nuclear Physics,
        Ulughbek, Tashkent 100214, Uzbekistan}
\altaffiltext{3}{International Centre for Theoretical Physics,
Strada Costiera 11, 34014 Trieste, Italy}




\begin{abstract}
\noindent

\noindent \noindent We study the structure of electromagnetic
field of slowly rotating magnetized star in a Randall-Sundrum II
type braneworld. The star is modeled as a sphere consisting of
perfect highly magnetized fluid with infinite conductivity and
frozen-in dipolar magnetic field. Maxwell's equations for the
external magnetic field of the star in the braneworld are
analytically solved in approximation of small distance from the
surface of the star.  We have also found numerical solution for
the electric field outside the rotating magnetized neutron star in
the braneworld in dependence on brane tension. The influence of
brane tension on the electromagnetic energy losses of the rotating
magnetized star is underlined. Obtained "brane" corrections are
shown to be relevant and have non-negligible values. In comparison
with astrophysical observations on pulsars spindown data they may
provide an evidence for the brane tension and, thus, serve as a
test for the braneworld model of the Universe.

\end{abstract}

\section{Introduction}

The study of magnetic and electric fields around the compact
objects is an important task for several reasons. First is that we
obtain information about such stars through their observable
characteristics, which are closely connected with electromagnetic
fields inside and outside the stars. Magnetic fields play an
important role in the life history of majority astrophysical
objects especially of compact relativistic stars which posses
surface magnetic fields of $10^{12}G$ and $\sim 10^{14}G$ in the
exceptional cases for magnetars. The strength of compact star's
magnetic field is one of the main quantities determining their
observability, for example as pulsars through magneto-dipolar
radiation. Electric field surrounding the star determines energy
losses from the star and therefore may be related with such
observable parameters as period of pulsar and it's time
derivation. The second reason is that we may test various theories
of gravitation through the study of compact objects for which
general relativity effects are especially strong. Considering
different metrics of space-time one may investigate the effect of
the different phenomena on evolution and behavior of stellar
interior and exterior magnetic fields. Then these models can be
checked through comparison of theoretical results with
observational data. The third reason may be seen in influence of
stellar magnetic and electric field on different physical
phenomena around the star, such as gravitational lensing and
motion of test particles.

In the Newtonian framework the exterior electromagnetic fields of
magnetized and rotating sphere are given in the classical paper of
\citet{d55} and interior fields are studied by many authors, for
example, in the paper of \citet{rt73}. In the general-relativistic
approach the study of the  magnetic field structure outside
magnetized  compact gravitational objects has been initiated by
the pioneering work of \citet{go64} and have been further extended
by number of authors (\citet{ac70}, \citet{p74}, \citet{ws83},
\citet{mh97}, \citet{mt92}, \citet{ram01a}, \citet{ram01b},
\citet{japan}), while in some papers (\citet{getal98},
\citet{pg97}, \citet{gpz00}, \citet{pgz00}, \citet{zr}) the work
has been completed by considering magnetic fields interior
relativistic star for the different models of stellar matter.
General-relativistic treatment for the structure of external and
internal stellar magnetic fields including numerical results has
shown that the magnetic field is amplified by the monopolar part
of gravitational field depending on the compactness of the
relativistic star.

We are interested in study of stellar electric and magnetic fields
in frames of recently popular model of the braneworld first
proposed in the work of \citet{RaSu99}. According to this model
our four-dimensional space-time is just a slice of
five-dimensional bulk and only gravity is the force which can
freely propagate between our space-time and bulk while other
fields are confined to four-dimensional Universe. The review of
braneworld models is given in the work of \citet{maar04} and some
cosmological and astrophysical implications of the braneworld
theories may be found in works \citet{maar00}, \citet{cs01},
\citet{lan01}, \citet{hm03}, \citet{ger06}, \citet{kg08},
\citet{MaMu05}. For astrophysical interests, static and
spherically symmetric exterior vacuum solutions of the braneworld
models were initially proposed by \citet{dmp00} which have the
mathematical form of the Reissner-Nordstr\"{0}m solution, in which
a tidal Weyl parameter $Q^\ast$ plays the role of the electric
charge squared of the general relativistic solution. It should be
noted that besides this solution was pioneering and there are many
different vacuum braneworld solutions at the moment, this solution
still stays interesting and actual and, for example, was recently
applied to the solar system tests in the paper of \citet{brhl10}.

Observational possibilities of testing the braneworld black hole
models at an astrophysical scale have been intensively discussed
in the literature during the last several years, for example,
through the gravitational lensing, the motion of test particles,
and the classical tests of general relativity (perihelion
precession, deflection of light, and the radar echo delay) in the
Solar System (see \citet{lobo08}). In the paper of \citet{pkh08}
the energy flux, the emission spectrum, and accretion efficiency
from the accretion disks around several classes of static and
rotating braneworld black holes have been obtained. The complete
set of analytical solutions of the geodesic equation of massive
test particles in higher dimensional spacetimes which can be
applied to braneworld models is provided in the recent paper of
\citet{Lam08}. The relativistic quantum interference effects in
the spacetime of slowly rotating object in braneworld and phase
shift effect of interfering particle in neutron interferometer
have been studied in the recent paper of \citet{mht10}. The
influence of the tidal charge onto profiled spectral lines
generated by radiating tori orbiting in the vicinity of a rotating
black hole has been studied in the paper of \citet{sstuch09}.
Authors showed that with lowering the negative tidal charge of the
black hole, the profiled line becomes flatter and wider, keeping
their standard character with flux stronger at the blue edge of
the profiled line. The role of the tidal charge in the orbital
resonance model of quasiperiodic oscillations in black hole
systems has been investigated in the paper of \citet{stuch09}. The
influence of the tidal charge parameter of the braneworld models
on some optical phenomena in rotating black hole spacetimes has
been extensively studied in the paper of \citet{ssstuch09}.

A braneworld corrections to the charged rotating black holes and
to the perturbations in the electromagnetic potential around black
holes are studied, for example, in works of \citet{ag05} and
\citet{aa10}. Our preceding paper, \citet{af08}, was devoted to
the stellar magnetic field configurations of relativistic stars in
dependence on brane tension. The present paper extends the paper
of \citet{af08} to the case of rotating relativistic star. In our
paper we will consider rotating spherically symmetric star in the
braneworld endowed with strong magnetic fields. We assume that the
star has dipolar magnetic field and the field energy is not strong
enough to affect the spacetime geometry, so we consider the
effects of the gravitational field of the star in the braneworld
on the magnetic and electric field structure without feedback. The
motion of test particles near black holes immersed in an
asymptotically uniform magnetic field and some gravity surrounding
structure, which provides the magnetic field has been intensively
studied in the paper of \citet{kon06}. The author has calculated
the binding energy for spinning particles on circular orbits. The
bound states of the massive scalar field around a rotating black
hole immersed in the asymptotically uniform magnetic field are
considered in the paper of \citet{kon07}. The uniform magnetic
field in the background of a five dimensional black hole has been
extensively studied in the work of \citet{alfr04}. In particular,
authors presented exact expressions for two forms of an
electromagnetic tensor and the electrostatic potential difference
between the event horizon of a five dimensional black hole and the
infinity.

The paper is organized as follows. In section \ref{meq} we present
a set of Maxwell's equations in the space-time of spherically
symmetric rotating relativistic compact star in the braneworld.
Section \ref{ss} is devoted for solutions of Maxwell's equations.
In subsection \ref{toy} we consider the solution for "toy model" -
monopolar structure of magnetic field of the star. This solution
is not realistic but it can be used to obtain first estimates of
the influence of brane tension on the electromagnetic field of the
star. In subsection \ref{srst} we are looking for analytical
solution of the Maxwell's equations for the exterior magnetic
field of the star. We obtain approximate solution of the
differential equation for magnetic field in the near vicinity of
the surface of the star. In subsection \ref{electr} we get the
differential equation for the electric field outside the star and
solve them numerically. We show that both magnetic and electric
fields will be essentially modified by five-dimensional gravity
effects. In subsection \ref{application} we investigate the
astrophysical application of obtained result, namely, calculate
energy losses from the slowly rotating magnetized neutron star in
the braneworld. The last section is devoted to the conclusions of
the research done.

Throughout, we use a space-like signature $(-,+,+,+)$ and a system
of units in which $G = 1 = c$. Greek indices are taken to run from
0 to 3 and Latin indices from 1 to 3; covariant derivatives are
denoted with a semi-colon and partial derivatives with a comma.

\section{Maxwell Equations In a Spacetime of Slowly Rotating Spherical Star
in the Braneworld}
\label{meq}

The metric for the space-time around rotating compact object in
the braneworld may be written in coordinates
${u,r,\theta,\varphi}$ as (see \citet{pkh08})
\begin{eqnarray}
&& ds^2=\Bigg[-(du+dr)^2+dr^2+\Sigma d\theta^2 \nonumber\\ &&
\qquad +(r^2+a^2)\sin^2\theta d\varphi^2 +2a\sin^2\theta dr
d\varphi \nonumber\\ && \qquad\qquad + G(du-a\sin^2\theta
d\varphi)^2\Bigg] \ ,
\end{eqnarray}
where $G=(2Mr-Q^*)/\Sigma$, $\Sigma=r^2+a^2\cos^2\theta$, $Q^*$ is
the is the bulk tidal charge, $M$ is the mass of the star,
parameter $a$ is related to the angular momentum of the star. In
this form the metric was presented in the paper of \citet{ag05}
and obtained in assumption that the axisymmetric and stationary
braneworld metric has the Kerr-Schild form and can be expressed in
the form of its linear approximation around the flat metric $ds^2
= \left(ds^2\right)_{flat} + H\left(l_{\mu}dx^{\mu}\right)^2$,
where $l_{\mu}$ is a null, geodesic vector field in both the flat
and full metrics, and $H$ is an arbitrary scalar function.

Applying the Boyer-Lindquist transformation
$du=dt-(r^2+a^2)dr/\Delta$, $d\varphi=d\phi-adr/\Delta$ with
$\Delta=r^2+a^2-2Mr+Q^*$ and assuming parameter $a$ to be small
and assuming parameter $a$ to be small one and assuming parameter
$a$ to be small one can obtain the exterior metric for slowly
rotating neutron star in the braneworld in the following form
\begin{eqnarray}
\label{metric} &&
ds^2=-A^2dt^2+H^2dr^2+r^2d\theta^2+r^2\sin^2\theta d\phi^2
\nonumber\\ && \qquad - 2\tilde{\omega}(r)r^2\sin^2\theta dt
d\phi\ ,
\end{eqnarray}
where
\begin{equation}
A^2(r)\equiv\left(1-\frac{2M}{r}+\frac{Q^*}{r^2}\right)=H^{-2}(r),\
\ \ \ \ \ \ \ r>R\ ,
\end{equation}
is the Reissner-Nordstr\"{o}m-type exact solution for the metric
outside the star, $\tilde{\omega}(r)=\omega(1-Q^*/2rM)$,
$\omega=2Ma/r^3$ is the angular velocity of the dragging of
inertial frames.

The general form of the first pair of general relativistic Maxwell
equations is given by
\begin{equation}
\label{maxwell_firstpair} 3! F_{[\alpha \beta, \gamma]} =  2
\left(F_{\alpha \beta, \gamma }
    + F_{\gamma \alpha, \beta} + F_{\beta \gamma,\alpha}
    \right) = 0 \ ,
\end{equation}
where $F_{\alpha \beta}$ is the electromagnetic field tensor.

The covariant components of the electromagnetic tensor are
explicitly expressed in terms of electric and magnetic field
components as
\begin{equation}
\label{fab_def} F_{\alpha\beta} \equiv 2 u_{[\alpha} E_{\beta]} +
    \eta_{\alpha\beta\gamma\delta}u^\gamma B^\delta \ ,
\end{equation}
where $u^{\alpha}$ is the four-velocity of observer, $T_{[\alpha
\beta]} \equiv \frac{1}{2}(T_{\alpha \beta} - T_{\beta \alpha})$ and
$\eta_{\alpha\beta\gamma\delta}$ is the pseudo-tensorial expression
for the Levi-Civita symbol $\epsilon_{\alpha \beta \gamma \delta}$
\begin{equation}
\label{levichivita}
\eta^{\alpha\beta\gamma\delta}=-\frac{1}{\sqrt{-g}}
    \epsilon_{\alpha\beta\gamma\delta} \ ,
    \quad
\eta_{\alpha\beta\gamma\delta}=
    \sqrt{-g}\epsilon_{\alpha\beta\gamma\delta} \ ,
\end{equation}

with $g\equiv {\rm det}|g_{\alpha\beta}|=-A^2H^2 r^4 \sin^2\theta$.

 The general form of the second pair of Maxwell
equations is given by
\begin{equation}
\label{maxwell_secondpair} F^{\alpha \beta}_{\ \ \ \ ;\beta} = 4\pi
J^{\alpha}
\end{equation}
where the four-current $J^{\alpha}$ is a sum of convection and
conduction currents
\begin{equation}
J^{\alpha}=\rho_e w^\alpha + j^\alpha \ ,
    \hskip 1.0cm j^\alpha w_\alpha \equiv 0\ ,
    \qquad
    j_\alpha = \sigma F_{ \alpha \beta}w^\beta \ ,
\end{equation}
where $\sigma$ is the electrical conductivity and $w^\alpha$ is
four-velocity of conducting medium as a whole.

The four-velocity of the 'zero angular momentum observer' (ZAMO)
in the metric \ref{metric} looks like
\begin{equation}
\label{uzamos} (u^{\alpha})_{_{\rm obs}}\equiv
    A^{-1}\bigg(1,0,0,\tilde{\omega}\bigg) \ ;
    \quad
(u_{\alpha})_{_{\rm obs}}\equiv
    A \bigg(- 1,0,0,0 \bigg) \ .
\end{equation}

Using the main equations (\ref{maxwell_firstpair}),
(\ref{maxwell_secondpair}) together with (\ref{fab_def}),
(\ref{levichivita}) and (\ref{uzamos}) one can obtain Maxwell
equations for slowly rotating neutron star in the braneworld for the
orthonormal reference frame as
\begin{eqnarray}
\label{max1a} &&\sin\theta \left(r^2B^{\hat r}\right)_{,r}+
    Hr\left(\sin\theta B^{\hat \theta}\right)_{,\theta} +
    H r B^{\hat \phi}_{\ , \phi} = 0 \ ,
\\
\label{max1b} &&\left({r\sin\theta}\right)\frac{\partial B^{\hat
r}}{\partial t}
     =  {A} \left[E^{\hat\theta}_{\ ,\phi}- \left(\sin\theta
    E^{\hat \phi} \right)_{,\theta}\right]\  \nonumber\\ && \qquad\qquad - (\tilde{\omega}r\sin\theta)B^{\hat r}_{,\phi} ,
\\
\label{max1c} &&\left({Hr\sin\theta}\right)
    \frac{\partial B^{\hat \theta}}{\partial t}
    = -AH E^{\hat r}_{\ ,\phi} +
    \sin\theta \left(r A E^{\hat \phi} \right)_{,r}  \nonumber\\ && \qquad\qquad  -(H\tilde{\omega}r\sin\theta)
    B^{\hat\theta}_{,\phi} \ ,
\\
\label{max1d} &&\left({Hr}\right)
    \frac{\partial B^{\hat \phi}}{\partial t}
    = - \left(r A E^{\hat \theta}\right)_{,r}
    + AH E^{\hat r}_{ \ ,\theta}  \nonumber\\ && \qquad\qquad  +\sin\theta(\tilde{\omega} r^2 B^{\hat r})_{,
    r} + H \tilde{\omega} r(\sin\theta B^{\hat\theta})_{,\theta}
\end{eqnarray}
\noindent and
\begin{eqnarray}
\label{max2a} &&\sin\theta\left(r^2 E^{\hat r} \right)_{,r}+
    {Hr}\left(\sin\theta E^{\hat \theta}\right)_{,\theta}
    + Hr E^{\hat \phi}_{\;,\phi}
    \nonumber \\ &&\qquad\qquad  =  {4\pi H}r^2\sin\theta J^{\hat t}\ ,
\\
\label{max2b} && A\left[\left(\sin\theta  B^{\hat \phi}
\right)_{,\theta}
    - B^{\hat\theta}_{\ ,\phi}\right] -
    (\tilde{\omega}r\sin\theta)E^{\hat r}_{,\phi}
     \nonumber \\ &&\qquad\qquad = \left({r\sin\theta}\right)
    \frac{\partial E^{\hat r}}{\partial t}
    +{4\pi}Ar\sin\theta J^{\hat r} \ ,
\\
\label{max2c} && AH B^{\hat r}_{\ ,\phi} - \sin\theta \left(r \ A
    B^{\hat \phi} \right)_{,r} -
    (H\tilde{\omega}r\sin\theta)E^{\hat\theta}_{,\phi}
     \nonumber \\ && \qquad\qquad =  \left({Hr\sin\theta}\right)
    \frac{\partial E^{\hat\theta}}{\partial t}
    +{4\pi AH}r\sin\theta J^{\hat\theta} \ ,
\\
\label{max2d} && \left(Ar B^{\hat \theta} \right)_{,r} - AH
    B^{\hat r}_{\ ,\theta} + \sin\theta(\tilde{\omega}r^2E^{\hat
    r})_{,r} \nonumber \\ && \qquad\qquad + H \tilde{\omega}r(\sin\theta E^{\hat\theta})_{,\theta}
     = \left({Hr}\right)
    \frac{\partial E^{\hat\phi}}{\partial t} \nonumber \\ &&
    \qquad\qquad\qquad
    +{4\pi AH}rJ^{\hat\phi} + 4\pi H \tilde{\omega}r^2\sin\theta J^{\hat t} \ .
\end{eqnarray}

\section{Stationary Solutions to Maxwell Equations}
\label{ss}

    We will look for stationary solutions of the Maxwell equation,
    i.e. for solutions in which we assume that the magnetic moment of the
star does not vary in time as a result of the infinite
conductivity of the stellar interior.

\subsection{A Special Monopolar Configuration for Magnetic Field of Slowly
Rotating Star in Braneworld} \label{toy}

First we consider the following magnetic field configuration as a
toy model
\begin{equation}
B^{\hat{r}} = B^{\hat{r}}(r) \neq 0 \ ,\qquad B^{\hat{\theta}} = 0\
.
\end{equation}

Although this form of magnetic field can not be considered
realistic, we will show that this toy model can be used to obtain
the first estimates of the influence of brane tension on the
electromagnetic field of the star. For this case, in the linear
approximation in the Lense-Thirring frequency $\omega$ Maxwell
equations (\ref{max1a}) and (\ref{max2d}) reduce to
\begin{equation}
\label{max1a_monopolar}  \left(r^2B^{\hat r}\right)_{,r} = 0 \ ,
\qquad B^{\hat r}_{,\theta} = 0 \ .
\end{equation}
The solution admitted by this equation is
\begin{equation}
\label{mf_monopolar} B^{\hat{r}} =\frac{\tilde{\mu}}{r^2}\  ,
\end{equation}
where $\tilde{\mu}$ is the integration constant being responsible
for the source of the monopolar magnetic field.

For the chosen configuration of the magnetic field due to the
infinite conductivity of the stellar matter one can easily find
the electric field as
\begin{equation}
E^{\hat{\theta}}   \neq 0 \ ,\qquad E^{\hat{r}} =E^{\hat{\phi}}=
0\ .
\end{equation}

Then, analytical solution of the Maxwell equation
\begin{equation}
(rAE^{\hat{\theta}})_{,r}-\tilde{\mu}\sin\theta(\tilde{\omega})_{,r}=0\
\end{equation}
has the following form
\begin{equation}
\label{Esol}
E^{\hat{\theta}}=\frac{\tilde{\omega}-C}{rA}\tilde{\mu}\sin\theta\
,
\end{equation}
where $C$ is the integration constant (see also \citet{aak08} for
the case of slowly rotating NUT star). To find $C$ we match
(\ref{Esol}) with the known inner solution for the  electric field
of the rotating star in the Newtonian spacetime. For the star
rotating with the angular frequency $\Omega$ we have (see
\citet{ra04})
\begin{equation}
E_{in}^{\hat{\theta}}=-\frac{\Omega r\sin\theta
B^{\hat{r}}_{in}}{N}\ .
\end{equation}

Remember now that the radial component of the magnetic field and
tangential components of the electric field are continuous at the
surface of the star and obtain (\ref{Esol}) as
\begin{equation}
\label{Esol1}
E^{\hat{\theta}}=\frac{\tilde{\omega}-\Omega}{rA}\tilde{\mu}\sin\theta\
.
\end{equation}

\begin{figure}
\includegraphics[width=0.47\textwidth]{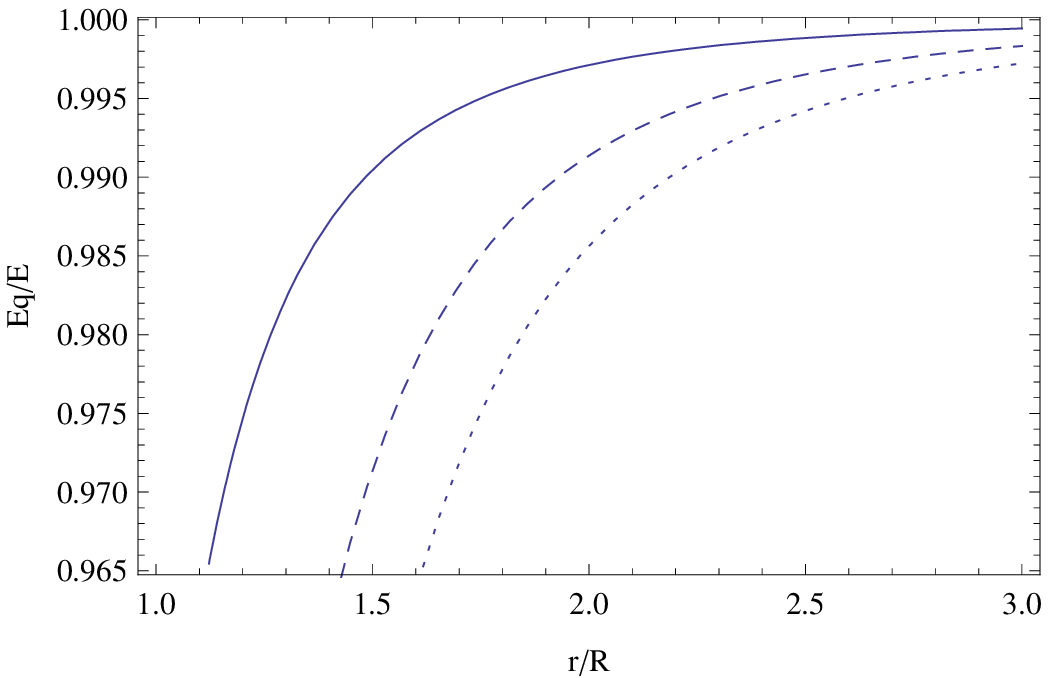}
\caption{Radial dependence of the ratio of electric field to that
when $q=0$ for the several values of $q=Q^*/M^2$: solid line
corresponds to $q=-1$, dashed to $q=-3$, dotted to $q=-5$.}
\label{monopolar}
\end{figure}

In the Fig.~\ref{monopolar} radial dependence of monopolar
$E/E_{q=0}$ is shown for the several values of parameter
$q=Q^*/M^2$ (we assume compactness parameter
$\varepsilon=M/R=1/3$). One can see when the star has monopolar
magnetic field the electric field is noticeably diminished due to
the presence of the brane tension.

\subsection{Stationary Solutions to Maxwell Equations for Dipolar
Magnetic Field of Slowly Rotating Magnetized Highly Conducting Star
in Braneworld} \label{srst}

Now we look for solution of the Maxwell equations
(\ref{max1a})--(\ref{max1d}) and (\ref{max2a})--(\ref{max2d})
assuming the realistic configurations of the magnetic field of the
star i.e. the dipolar one.

We look for separable solutions of Maxwell equations in the form
\begin{eqnarray}
\label{ansatz_1} && B^{\hat r}(r,\theta) = F(r)\cos\theta\ , \quad
B^{\hat \theta}(r,\theta) = G(r)\sin\theta\ , \nonumber \\ &&
\qquad\qquad\quad B^{\hat \phi}(r,\theta) = 0\ ,
\end{eqnarray}
assuming that magnetic field of the star is dipolar and functions
$F(r)$ and $G(r)$ will account for the relativistic corrections
due to a curved background spacetime.

Maxwell equations (\ref{max1a}), (\ref{max2b})--(\ref{max2d}) with
the ansatz (\ref{ansatz_1}), yield the following set of equations
\begin{eqnarray}
\label{ir_1} &&\left(r^2 F\right)_{, r} + 2 Hr G = 0 \ ,
\\\nonumber\\
\label{ir_2} &&\left(r A G\right)_{, r}    +  F = 0\ .
\end{eqnarray}

The exterior solution for the magnetic field is simplified by the
knowledge of explicit analytic expressions for the metric
functions $A$ and $H$. In particular, after defining $A =H^{-1}=
(1 - 2M/r+Q^*/r^2)^{1/2}$, the system (\ref{ir_1})--(\ref{ir_2})
can be written as a single, second-order ordinary differential
equation for the unknown function $F$

\begin{equation}
\label{leg_eq_sim}
\frac{d}{dr}\left[\left(1-\frac{2M}{r}+\frac{Q^*}{r^2}\right)
    \frac{d}{dr}\left(r^2 F\right)\right] - 2F = 0 \ .
\end{equation}

The analytical solution of equation (\ref{leg_eq_sim}) exists in
the literature when the parameter $Q^*=0$ (see, for example,
\citet{ram01a}, \citet{ram01b}). In the case $Q^*\neq0$ the
equation is more complicated but it allows analytical solution
under simplification near the surface of the star. First we
rewrite equation (\ref{leg_eq_sim}) as
\begin{eqnarray}
\label{f1} &&
r^2\left(1-\frac{2M}{r}+\frac{Q^*}{r^2}\right)\frac{d^2F}{dr^2}
\nonumber
\\ && \qquad +4r\left(1-\frac{3M}{2r}+
\frac{Q^*}{2r^2}\right)\frac{dF}{dr} -\frac{2Q^*}{r^2}F=0
\end{eqnarray}
and introduce the new variable $z$ so that $r=R+\delta=R(1+z)$,
$z=\delta/R$. Using also compactness parameter $\varepsilon=M/R$
and introducing $q=Q^*/M^2$ we get

\begin{eqnarray}
\label{f2}
&&\left[(1+z)^4-2\varepsilon(1+z)^3+q\varepsilon^2(1+z)^2\right]\frac{d^2F}{dz^2}
\nonumber
\\ && \qquad +\left[4(1+z)^3-6\varepsilon(1+z)^2+2q\varepsilon^2(1+z)\right]\frac{dF}{dz}
\nonumber \\ && \qquad\qquad -2q\varepsilon^2F=0\ .
\end{eqnarray}

Now one may consider the region just above the surface of the
star, where $z\ll1$, expand the coefficients standing before the
derivatives into series and leave only terms of first order of $z$
and get

\begin{eqnarray}
\label{f3}
&&\left[(2q\varepsilon^2-6\varepsilon+4)z+(q\varepsilon^2-2\varepsilon+1)\right]\frac{d^2F}{dz^2}
\nonumber
\\ && \qquad +\left[(2q\varepsilon^2-12\varepsilon+12)z+(2q\varepsilon^2-6\varepsilon+4)\right]\frac{dF}{dz}
\nonumber \\ && \qquad\qquad -2q\varepsilon^2F=0\ .
\end{eqnarray}

Obtained equation is the particular case of equation (see
\citet{kamke})
\begin{eqnarray}
&& (a_2x+b_2)f''+(a_1x+b_1)f'+(a_0x+b_0)f=0, \nonumber \\ &&
\qquad\qquad |a_2|+|b_2|>0\ ,
\end{eqnarray}
which in the case when $a_2\neq0$ with the help of substitutions
$f(x)=e^{kx}\eta(\xi)$, $a_2\xi=a_2x+b_2$, where $k$ is the root
of the square equation $a_2k^2+a_1k+a_0=0$ may be rewritten in the
following form
\begin{eqnarray}
&&
a_2\xi\eta''+\left[(2sa_2+a_1)\xi+\frac{a_2b_1-a_1b_2}{a_2}\right]\eta'
\nonumber \\ && \qquad +
\left(\frac{a_2b_0-a_0b_2}{a_2}+\frac{a_2b_1-a_1b_2}{a_2}s\right)\eta=0\
.
\end{eqnarray}

In the case of our equation (\ref{f3}) $a_0=0$, so $s=0$ will be
one of the roots of square equation. This will make the
substitutions to be $F(z)=\eta(\xi)$, $\xi=z+b_2/a_2$, where

\begin{eqnarray}
a_2&=&(2q\varepsilon^2-6\varepsilon+4)\ ,\qquad
b_2=(q\varepsilon^2-2\varepsilon+1)\
,\nonumber\\
a_1&=&(2q\varepsilon^2-12\varepsilon+12)\ , \qquad
b_1=(2q\varepsilon^2-6\varepsilon+4)\ , \nonumber \\
b_0&=&-2q\varepsilon^2\ ,
\end{eqnarray}
so equation (\ref{f3}) may be rewritten as

\begin{equation}
\label{hyper} \xi\eta''+(a\xi+b)\eta'+(c\xi+d)\eta=0
\end{equation}

with
\begin{eqnarray}
&&
a=\frac{\varepsilon^2q-6\varepsilon+6}{\varepsilon^2q-3\varepsilon+2}\
, \nonumber \\ &&
b=\frac{\varepsilon^4q^2-4\varepsilon^3q+\varepsilon^2(6+q)-6\varepsilon+2}
{2(\varepsilon^2q-3\varepsilon+2)^2}\ ,\qquad \nonumber
\\ &&   c=0\ ,\qquad
d=-\frac{\varepsilon^2q}{\varepsilon^2q-3\varepsilon+2}\ .
\end{eqnarray}

Equation (\ref{hyper}) belongs to type of degenerate
hypergeometric equation. Solution of this equation have the
following sign (\citet{kamke})

\begin{eqnarray}
\label{sol1} &&
\eta=\xi^{-\frac{b}{2}}e^{-\frac{a\xi}{2}}\Bigg[C_1(a\xi)^{\frac{b}{2}}e^{-\frac{a\xi}{2}}
{_1F_1}\left(b-\frac{d}{a},b,a\xi\right)\nonumber
\\ && \quad
+C_2(a\xi)^{1-\frac{b}{2}}e^{-\frac{a\xi}{2}}
{_1F_1}\left(1-\frac{d}{a},2-b,a\xi\right)\Bigg]\ ,
\end{eqnarray}
where the constants $C_1$ and $C_2$ can be determined from the
boundary conditions and definition
\begin{equation}
\label{series}
_1F_1(l,m,x)=1+\sum^\infty_{n=1}\frac{l(l+1)...(l+n-1)x^n}{m(m+1)...(m+n-1)n!}\
.
\end{equation}

Turning back to $F(z)$ from $\eta(\xi)$, one can get the radial
eigenfunction of magnetic field outside the slowly rotating star
in the braneworld as

\begin{eqnarray}
\label{FQfinal} && F(z)=\left(z+s\right)^{-\frac{b}{2}}
e^{-a\left(z+s\right)} \nonumber \\ && \qquad \times
\Bigg\{C_1\left[a\left(z+s\right)\right]^{\frac{b}{2}}
{_1F_1}\left(b-\frac{d}{a},b,a(z+s)\right) \nonumber\\ &&
\qquad\qquad +C_2\left[a\left(z+s\right)\right]^{1-\frac{b}{2}}
\nonumber\\ && \qquad\qquad\quad \times
{_1F_1}\left(1-\frac{d}{a},2-b,a(z+s)\right)\Bigg\}\ ,
\end{eqnarray}
where

\begin{equation}
s=\frac{\varepsilon^2q-2\varepsilon+1}{2\varepsilon^2q-6\varepsilon+4}\
.
\end{equation}

The integration constants $C_1$ and $C_2$ may be found with the
help of exact solution for the case $q=0$, which has the following
form (see e.g. \citet{ram01a})
\begin{equation}
F_{exact}(r)=-\frac{3}{4M^3}\left[\ln{N^2}+\frac{2M}{r}\left(1+\frac{M}{r}\right)\right]\mu
\end{equation}
and using variable $z$ one can obtain it as
\begin{eqnarray}
&&
F_{exact}(z)=-\frac{3}{4M^3}\Bigg[\ln{\left(1-\frac{2}{z+\varepsilon}\right)}
\nonumber \\ && \qquad\qquad
+\frac{2}{z+\varepsilon}\left(1+\frac{1}{z+\varepsilon}\right)\Bigg]\mu\
,
\end{eqnarray}
where $\mu$ is dipolar moment.

Using conditions
\begin{eqnarray}
&& F(z)_{q=0}|_{z=0}=F_{exact}(z)|_{z=0}\ ,\nonumber \\
&& \frac{dF(z)}{dz}_{q=0}|_{z=0}=\frac{dF_{exact}(z)}{dz}|_{z=0}
\end{eqnarray}
and assuming $\varepsilon=1/3$ we numerically found
$C_1\approx0.28\mu/M^3$, $C_2\approx-0.25\mu/M^3$.

The function $G(z)$ may be found through the relation
\begin{equation}
\label{G} G(r)=-\frac{1}{2r}A(r)(r^2F)_{,r}\ .
\end{equation}

\begin{figure}
\includegraphics[width=0.47\textwidth]{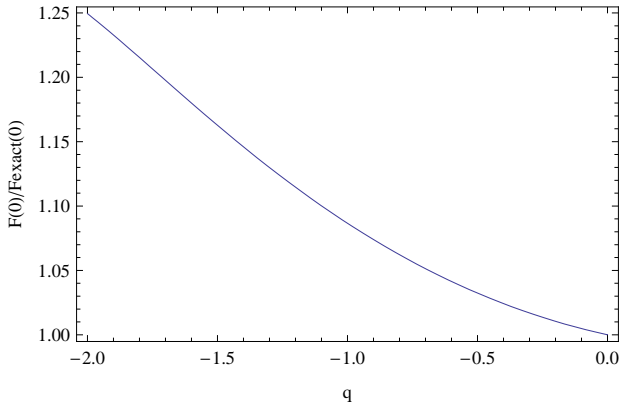}
\caption{$F(z)|_{z=0}/F_{exact}|_{z=0}$ as a function of
$q=Q^*/M^2$.} \label{Fgraph}
\end{figure}

\begin{figure}
\includegraphics[width=0.47\textwidth]{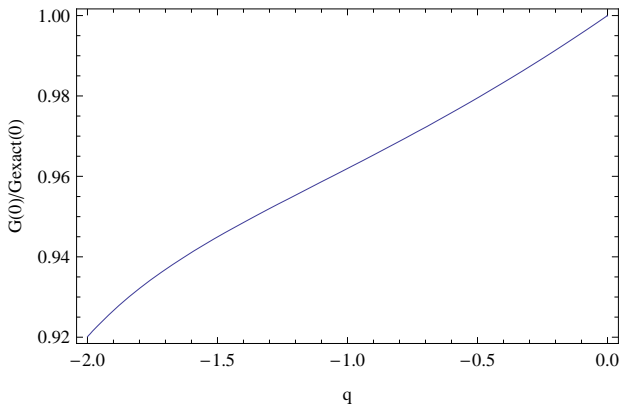}
\caption{$G(z)|_{z=0}/G_{exact}|_{z=0}$ as a function of
$q=Q^*/M^2$.} \label{Ggraph}
\end{figure}

$F(z)$ and $G(z)$ as functions of $q$ are presented in the figures
\ref{Fgraph} and \ref{Ggraph}, correspondingly. Functions are
taken at the surface of the star and normed on its exact values
for the case $Q^*=0$. It is seen from the graphs that the value of
the surface magnetic field is noticeably modified due to the
presence of brane tension $Q^*$, especially the radial component
of $B$. The radial component $B^{\hat{r}}(R,\theta)$ increases
with the growth of $|q|$ while the angular component
$B^{\hat{\theta}}(R,\theta)$ decreases. It should be noted that
the surface value of magnetic field is of great importance and has
strong influence on conditions of emission generation and energy
losses from the rotating magnetized star.


\subsection{Solution for Electric Field of Slowly Rotating Magnetized Highly Conducting Star
in Braneworld} \label{electr}

The search of solution for the electric field is  more complicated
with compare to that for the magnetic field. To simplify our task
we will use the exact solution for magnetic field, obtained in the
work of \citet{ram01a} (for the case $Q^*=0$), and, therefore,
consider first order corrections by angular velocity and brane
tension. With the help of exact expressions for the stationary
magnetic field external to a misaligned magnetized relativistic
star:
\begin{eqnarray}
&& B^{\hat{r}}=-\frac{3}{4M^3}\left[\ln
N^2+\frac{2M}{r}\left(1+\frac{M}{r}\right)\right]\nonumber \\ &&
\qquad\qquad
\times(\cos\chi\cos\theta+\sin\chi\sin\theta\cos\lambda)\mu\ , \\
&& B^{\hat{\theta}}=\frac{3N}{4M^2 r}\left[\frac{r}{M}\ln N^2
+\frac{1}{N^2}+1\right]\nonumber \\ && \qquad\qquad
\times(\cos\chi\sin\theta-\sin\chi\cos\theta\cos\lambda)\mu\ , \\
&& B^{\hat{\theta}}=\frac{3N}{4M^2 r}\Bigg[\frac{r}{M}\ln N^2
\nonumber \\ && \qquad\qquad
+\frac{1}{N^2}+1\Bigg](\sin\chi\sin\lambda)\mu
\end{eqnarray}
one can obtain Maxwell equations for the electric field around
misaligned star in the braneworld as
\begin{eqnarray}
\label{MaxEl} && A\left[(\sin\theta
E^{\hat{\phi}})_{,\theta}-E^{\hat{\theta}}_{,\phi}\right]
=\frac{3\bar{\omega}r}{4 M^{3}}\mu\Bigg[\ln N^2\nonumber
\\ && \qquad\qquad +\frac{2M}{r}\left(1+\frac{M}{r}\right)\Bigg]\sin\chi\sin^2\theta\sin\lambda\
, \\
&& E^{\hat{r}}_{,\phi}-\sin\theta(rAE^{\hat{\phi}})_{,r}
=\frac{N}{A}\frac{3\bar{\omega}}{4 M^2}\mu\Bigg[\frac{r}{M}\ln N^2
\nonumber
\\ && \qquad\qquad +\frac{1}{N^2}+1\Bigg]\sin\chi\sin\theta\cos\theta\sin\lambda\ ,
 \\
&& (rAE^{\hat{\theta}})_{,r}-E^{\hat{r}}_{,\theta}=
\frac{3\tilde{\omega}}{2
M^2}\left(\frac{N}{A}-1\right)\mu\Bigg[\frac{r}{M}\ln N^2
\nonumber \\
&& \qquad +\frac{1}{N^2}+1\Bigg]\sin\theta(\cos\theta\cos\chi
+\sin\chi\cos\lambda\sin\theta)\nonumber \\ && \qquad
+\frac{N}{A}\frac{3\bar{\omega}}{4 M^2}\mu\left[\frac{r}{M}\ln N^2
+\frac{1}{N^2}+1\right]\sin\chi\cos\lambda \nonumber
\\ && \qquad + \frac{3\tilde{\omega}}{4
M^3}\mu\frac{\left(1-\frac{2Q^*}{3rM}\right)}{\left(1-\frac{Q^*}{2rM}\right)}\left[\ln
N^2+\frac{2M}{r}\left(1+\frac{M}{r}\right)\right]\nonumber \\ &&
\qquad \times \sin\theta(\cos\theta\cos\chi
 +\sin\chi\cos\lambda\sin\theta)\ , \\
&& A\sin\theta(r^2E^{\hat{r}})_{,r}+r(\sin\theta
E^{\hat{\theta}})_{,\theta}+rE^{\hat{\phi}}_{,\phi}=0\ .
\end{eqnarray}

The form of the external electromagnetic field around a misaligned
rotating magnetized sphere was found in the paper of \citet{d55}.
Taking into account these solutions we look for the simplest
solutions for the equations (\ref{MaxEl}) in the following form
\begin{eqnarray}
\label{commonE} &&
E^{\hat{r}}=(f_1+f_3)\cos\chi(3\cos^2\theta-1)\nonumber \\ &&
\qquad\qquad +(g_1+g_3)3\sin\chi\cos\lambda\sin\theta\cos\theta\ ,
\\ &&
E^{\hat{\theta}}=(f_2+f_4)\cos\chi\sin\theta\cos\theta+(g_2+g_4)\sin\chi\cos\lambda
\nonumber
\\ && \qquad\qquad  -(g_5+g_6)(\cos^2\theta-\sin^2\theta)\sin\chi\cos\lambda\ ,
\\ && E^{\hat{\phi}}=[g_5+g_6-(g_2+g_4)]\sin\chi\cos\theta\sin\lambda\
,
\end{eqnarray}
where the eigenfunctions $f_1-f_4$, $g_1-g_6$ have the radial
dependence only. Using Maxwell equations (\ref{MaxEl}) one can
find the following set of linear differential equations for these
functions:

\begin{eqnarray}
\label{comf}
&& A(r^2f_1)_{,r}+rf_2=0\ , \\
&& (rAf_2)_{,r}+6f_1=0\ , \\
&& A(r^2f_3)_{,r}+rf_4=0\ , \\
&& (rAf_4)_{,r}+6f_3-
\frac{9\tilde{\omega}r}{4M^3}\mu\frac{\left(1-\frac{2Q^*}{3rM}\right)}{\left(1-\frac{Q^*}{2rM}\right)}\nonumber
\\ && \  \times \left[\ln
N^2+\frac{2M}{r}\left(1+\frac{M}{r}\right)\right] \nonumber
\\ && \   -\frac{6\tilde{\omega}}{4M^2}\mu\left(\frac{N}{A}-1\right)
\left[\frac{r}{M}\ln N^2 +\frac{1}{N^2}+1\right]=0\ , \\
&& A(r^2g_1)_{,r}+2rg_5=0\ , \\
&& (rAf_5)_{,r}+3g_1=0\ , \\
&& A(r^2g_3)_{,r}+2rg_6=0\ , \\
&& (rAg_6)_{,r}+3g_3-
\frac{9\tilde{\omega}r}{8M^3}\mu\frac{\left(1-\frac{2Q^*}{3rM}\right)}{\left(1-\frac{Q^*}{2rM}\right)}
\nonumber \\ && \  \times \left[\ln
N^2+\frac{2M}{r}\left(1+\frac{M}{r}\right)\right] \nonumber
\\ && \   -\frac{6\tilde{\omega}}{8M^2}\mu\left(\frac{N}{A}-1\right)
\left[\frac{r}{M}\ln N^2 +\frac{1}{N^2}+1\right]=0\ .
\end{eqnarray}

Combining properly these equations one can obtain the following
differential equations of second order for unknown functions $f_1$
and $f_3$ as
\begin{eqnarray}
\label{f13} &&
\frac{d}{dr}\left[\left(1-\frac{2M}{r}+\frac{Q^*}{r^2}\right)\frac{d}{dr}(r^2
f_1)\right]-6f_1=0 \ , \\
&&
\frac{d}{dr}\left[\left(1-\frac{2M}{r}+\frac{Q^*}{r^2}\right)\frac{d}{dr}(r^2
f_3)\right]-6f_3
\\ && \  +\frac{9\tilde{\omega}r}{4M^3}\mu\frac{\left(1-\frac{2Q^*}{3rM}\right)}{\left(1-\frac{Q^*}{2rM}\right)}
\left[\ln N^2+\frac{2M}{r}\left(1+\frac{M}{r}\right)\right]
\nonumber
\\ && \
+\frac{6\tilde{\omega}}{4M^2}\mu\left(\frac{N}{A}-1\right)
\left[\frac{r}{M}\ln N^2 +\frac{1}{N^2}+1\right]=0\ .
\end{eqnarray}

From the system of equations (\ref{comf}) one can notice that
functions $f$ and $g$ are connected with the following relations
\begin{equation}
g_1=f_1\ ,\quad g_3=f_3\ ,\quad g_5=\frac{f_2}{2}\ ,\quad
g_6=\frac{f_4}{2}\ .
\end{equation}

Functions $g_2$ and $g_4$ can be found directly from the system of
Maxwell equations (\ref{MaxEl}), using (\ref{commonE}), as
\begin{eqnarray}
g_2=\frac{3\Omega r}{8M^3cA}\left[\ln
N^2+\frac{2M}{r}\left(1+\frac{M}{r}\right)\right]\mu\ , \\
g_4=-\frac{\tilde{\omega}}{\Omega}g_2=-\frac{3\tilde{\omega}
r}{8M^3cA}\left[\ln
N^2+\frac{2M}{r}\left(1+\frac{M}{r}\right)\right]\mu\ .
\end{eqnarray}

\begin{figure}
\includegraphics[width=0.47\textwidth]{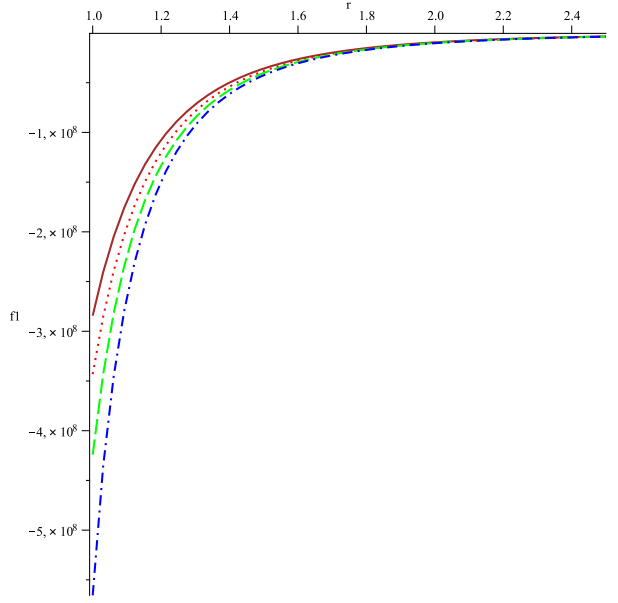}
\caption{The behavior of the radial electric field eigenfunction
$f_1$ for the different values of the brane tension (solid line
corresponds to $q=0$, dotted to $q=-1$, dashed to $q=-2$,
dashdotted to $q=-3$).} \label{elfield}
\end{figure}

We are looking for numerical solutions of the second order ordinary
differential equation (\ref{f13}) with the help of the Runge-Kutta
fifth order method. To do this we assume the solution to be
asymptotically Newtonian, namely
\begin{equation}
E^{\hat{r}}_{Newt}=-\frac{\mu\Omega R^2}{c r^4}\ ,
\end{equation}
since in the $r\rightarrow\infty$ limit $Q^*/r^2$ and $M/r$ are
negligibly small and do not give any contribution to the magnetic
field. As initial conditions to solve ODE we have taken the values
of the Newtonian expression and its derivation on a some large
radii. With such a prescription, the equation is integrated
inwards up to the surface of the relativistic star. Corresponding
graphs are presented in the figure \ref{elfield} for the different
values of brane tension $q=Q^*/M^2$.

As it is seen from the figure, the modification of electric field
for the different values of $q$ has the magnitude of the order of
tens percents of the standard value for the case when $q=0$.

\subsection{Astrophysical Consequences} \label{application}

Assume that the oblique rotating magnetized star is observed as
radio pulsar through magnetic dipole radiation. Then the
luminosity of the relativistic star in the case of a purely
dipolar radiation, and the power radiated in the form of dipolar
electromagnetic radiation, is given by \citet{ra04}
\begin{equation}
\label{dipole_energy_loss} L_{em} = \frac{\Omega^4 R^6 B^2_R}{6
c^3}\sin^2\chi
    \ ,
\end{equation}
where subscript $R$ denotes the value at $r=R$.

Considering slowly rotating magnetized neutron star in the
braneworld model one can see that the general relativistic
braneworld corrections emerging in expression
(\ref{dipole_energy_loss}) will be partly due to the magnetic
field amplification at the stellar surface and partly to the
increase in the effective rotational angular velocity produced by
the gravitational redshift as $\Omega_{_Q^*} = \Omega/A_{_R}$.

\begin{figure}
\includegraphics[width=0.47\textwidth]{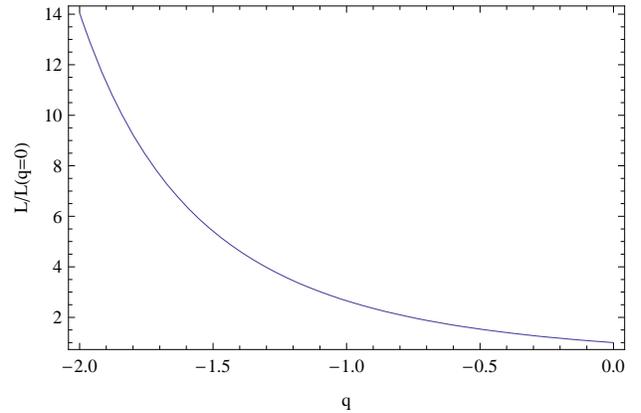}
\caption{The ration $\frac{L_{em\ q\neq0}}{L_{em\ q=0}}$ as a
function of $q=Q^*/M^2$.} \label{Losses}
\end{figure}

The presence of a braneworld tension has the effect of enhancing
the rate of energy loss through dipolar electromagnetic radiation
by an amount which can be easily estimated to be
\begin{equation}
\label{dipole_energy_loss_cf} \frac{L_{em\ q\neq0}}{L_{em\ q=0}}=
\left(\frac{F_R}{F_{exact\
R}}\right)^2\left(\frac{N_R}{A_R}\right)^4\ ,
\end{equation}
whose dependence from $q$ is shown in Fig. (\ref{Losses}) and
which may reach significant values depending on the value of the
brane tension parameter.

Noting that the energy losses of the pulsar may be related with
the frequency of its rotation and the derivative of this frequency
by time as
\begin{equation}
L_{em}=-\tilde{I}(\Omega\dot{\Omega})
\end{equation}
where $\tilde{I}$ is the general relativistic moment of inertia of
the star (see, for example~\citet{mak}),
\begin{equation}
\tilde{I}\equiv\int d^3\mathbf{x}\frac{\sqrt{\gamma}}{N}\rho
r^2\sin^2\theta\ ,
\end{equation}
where $\rho(r)$ is the total energy density, $\gamma$ is the
determinant of the three metric of slowly rotating star and
$d^3\mathbf{x}$ is the coordinate volume element, one may find the
following relation for the time derivatives of rotation period of
the pulsar
\begin{equation}
\frac{\dot{P}_{q\neq0}}{\dot{P}_{q=0}}=
\frac{\tilde{I}}{\tilde{I}_{q}}\frac{N_R}{A_R}\left(\frac{F_R}{F_{exact
R}}\right)^2\ ,
\end{equation}
where
\begin{equation}
\tilde{I}_{q}\equiv\int d^3\mathbf{x}\frac{\sqrt{\gamma_q}}{A}\rho
r^2\sin^2\theta\ ,
\end{equation}
and $\gamma_q$ is the determinant of the three metric of slowly
rotating neutron star in the braneworld determined by equation
(\ref{metric}).

Expression (\ref{dipole_energy_loss})  could be used to
investigate the rotational evolution of magnetized neutron stars
with predominant dipolar magnetic field anchored in the crust
which converting its rotational energy into electromagnetic
radiation. It should be also noted, that detailed investigation of
general relativistic effects for Schwarzschild stars has already
been performed by \citet{pgz00}, who have paid special attention
to the general relativistic corrections that need to be included
for a correct modeling of the thermal evolution but also of the
magnetic and rotational evolution.

\section{Conclusions}

In  the present paper we considered modifications of
electromagnetic field of a rotating magnetized neutron stars in
the braneworld. We formulated Maxwell's equations for the case of
slowly rotating magnetized compact star with non-zero brane
tension and considered the particular case of monopolar magnetic
field. Despite this configuration may not be considered as
realistic it helped us to see the strong influence of brane
tension on the electric field of the gravitating object. As the
analytical solution is always more valuable for further
calculations we attempted to find analytical solution for the
dipolar magnetic field configuration. We have derived an
approximate analytical expression for the magnetic field just near
the surface of the star as a solution of II type hypergeometric
equation. This region of the magnetosphere is very important
because exactly in this region the processes of plasma generation
responsible for the radio emission take place.

We have got equations for the electric field of the slowly
rotating magnetized neutron star on branes and solved one of them
numerically for different values of brane tension. It is shown
that the effect of brane tension on the electromagnetic field of
the star is non-negligible (may have the order of tens percents of
the initial value) and may help in future in testing the
braneworld model.

As an important application of the obtained results we have
calculated energy losses of slowly rotating magnetized neutron
star in the braneworld and found that the star with non-zero brane
parameter will lose more energy than typical rotating neutron star
in general relativity. The obtained dependence may be combined
with the astrophysical data on pulsar period slowdowns and be
useful in further investigations of the possible
detection/estimation of the brane parameter.

\section*{Acknowledgments}

This research is supported in part by the UzFFR (projects 1-10 and
11-10) and projects FA-F2-F079, FA-F2-F061 of the UzAS and by the
ICTP through the OEA-PRJ-29 project.



\begin{thebibliography}{Aliev \& G\"{u}mr\"{u}k\c{c}\"{u}o\v{g}lu(2005)}

\bibitem[Abdujabbarov et.al.(2008)]{aak08}{{Abdujabbarov}, A.~ A., {Ahmedov}, B.~J. and {Kagramanova}, V.~G.} 2008,
Gen. Rel. Grav., 40, {2515-2532}

\bibitem[Abdujabbarov \& Ahmedov(2010)]{aa10}{{Abdujabbarov}, A.~A. and {Ahmedov}, B.~J.} 2010, Phys. Rev. D, 81, 044022

\bibitem[Ahmedov \& Fattoyev(2008)]{af08}{{Ahmedov}, B.~J. and {Fattoyev}, F.~J.} 2008, Phys. Rev. D, 78, 047501

\bibitem[Aliev \& Frolov(2004)]{alfr04}{{Aliev}, A.~N. and {Frolov}, V.~P.} 2004, Phys. Rev. D, 69, 084022

\bibitem[Aliev \& G\"{u}mr\"{u}k\c{c}\"{u}o\v{g}lu(2005)]{ag05}{{Aliev}, A.~N. and
{G\"{u}mr\"{u}k\c{c}\"{u}o\v{g}lu}, A.~E.} 2005, Phys. Rev. D, 71,
104027

\bibitem[Anderson \& Cohen(1970)]{ac70}{{Anderson}, J.~L. and {Cohen}, J.~M.} 1970, Astrophys.
        Space Science, 9, {146-152}

\bibitem[B\"{o}hmer et.al.(2010)]{brhl10}{{B\"{o}hmer}, C.~G., {De Risi}, G., {Harko}, T. and {Lobo},
F.~S.~N.} 2010, Class. Quantum Grav., 27, 185013

\bibitem[B\"{o}hmer et.al.(2008)]{lobo08}{{B\"{o}hmer}, C.~G., {Harko}, T. and {Lobo}, F.~S.~N.} 2008,
Class. Quantum Grav., 25, 045015

\bibitem[Campos \& Sopuerta(2001)]{cs01}{{Campos}, A. and {Sopuerta}, C.~F.} 2001, Phys. Rev. D, 63, 104012

\bibitem[Dadhich et.al.(2000)]{dmp00}{{Dadhich}, N.~K., {Maartens}, R., {Papodopoulos}, P. and {Rezania}, V.} 2000,
Phys. Lett. B, 487, {1-6}

\bibitem[Deutsch(1955)]{d55}{{Deutsch}, A.~J.} 1955, Ann. Astrophys. 1,
{1-10}

\bibitem[Geppert et.al.(2000)]{gpz00}{{Geppert}, U., {Page}, D. and {Zannias}, T.} 2000, Phys. Rev. D, 61, 123004

\bibitem[Gergely(2006)]{ger06}{{Gergely}, L.~A.} 2006, Phys. Rev. D, 74, 024002

\bibitem[Ginzburg \& Ozernoy(1964)]{go64}{{Ginzburg}, V.~L. and {Ozernoy}, L.~M.} 1964, Zh. Eksp. Teor.
        Fiz., 47, {1030-1040}

\bibitem[Gupta et.al.(1998)]{getal98}{{Gupta}, A., {Mishra}, A., {Mishra}, H. and {Prasanna}, A.~R.} 1998,
Class. Quantum Grav., 15, {3131-3145}

\bibitem[Hackmann et.al.(2008)]{Lam08}{{Hackmann}, E., {Kagramanova}, V., {Kunz}, J. and {L\"{a}mmerzahl}, C.} 2008,
Phys. Rev. D, 78, 124018

\bibitem[Harko \& Mak(2003)]{hm03}{{Harko}, T. and {Mak}, M.~K.} 2003, Class. Quantum Grav., 20,
{407-422}

\bibitem[Kamke(1959)]{kamke}{{Kamke}, E.} Differential Gleichungen L\"{o}sungsmethoden und
L\"{o}sungen. Leipzig (1959)

\bibitem[Kojima et.al.(2004)]{japan}{{Kojima}, Y., {Matsunaga}, N. and {Okita}, T.} 2004, {\mnras}, 348,
{1388-1394}

\bibitem[Konoplya(2006)]{kon06}{{Konoplya}, R.~A.} 2006, Phys. Rev. D, 74, 124015

\bibitem[Konoplya(2007)]{kon07}{{Konoplya}, R.~A.} 2007, Phys. Lett. B, 644,
{219-223}

\bibitem[Kovacs \& Gergely(2008)]{kg08}{{Kovacs}, Z. and {Gergely}, L.~A.} 2008, Phys. Rev. D, 77, 024003

\bibitem[Langlois(2001)]{lan01}{{Langlois}, D.} 2001, Phys. Rev. Lett., 86,
{2212-2215}

\bibitem[Maartens(2000)]{maar00}{{Maartens}, R.} 2000, Phys. Rev. D, 62, 084023

\bibitem[Maartens(2004)]{maar04}{{Maartens}, R.} 2004, Living Rev. Rel., 7,
{1-58}

\bibitem[Majumdar \& Mukherjee(2005)]{MaMu05}{{Majumdar}, A.~S. and {Mukherjee}, N.} 2005, Int. J. Mod. Phys. D, 14,
{1095-1129}

\bibitem[Mamadjanov et.al.(2010)]{mht10}{{Mamadjanov}, A.~I., {Hakimov}, A.~A. and {Tojiev}, S.~R.} 2010,
Mod. Phys. Lett. A, 25, 4, {243-256}

\bibitem[Morozova et.al.(2008)]{mak}{{Morozova}, V.~S., {Ahmedov}, B.~J. and {Kagramanova}, V.~G.} 2008,
{\apj}, 684, {1359-1365}

\bibitem[Muslimov \& Tsygan(1992)]{mt92}{{Muslimov}, A. and {Tsygan}, A.~I.} 1992, {\mnras}, 255,
{61-70}

\bibitem[Muslimov \& Harding(1997)]{mh97}{{Muslimov}, A. and {Harding}, A.~K.} 1997, {\apj}, 485,
{735-746}

\bibitem[Page et.al.(2000)]{pgz00}{{Page}, D., {Geppert}, U. and {Zannias}, T.} 2000, Astron. Astrophys., 360,
{1052-1066}

\bibitem[Petterson(1974)]{p74}{{Petterson}, J.~A.} 1974, Phys. Rev. D, 10,
{3166-3170}

\bibitem[Prasanna \& Gupta(1997)]{pg97}{{Prasanna}, A.~R. and {Gupta}, A.} 1997, Nuovo Cim. B, 112,
{1089-1106}

\bibitem[Pun et.al.(2008)]{pkh08}{{Pun}, C.~S.~J., {Kov\'{a}cs}, Z. and {Harko}, T.} 2008, Phys. Rev. D, 78,
084015

\bibitem[Randall \& Sundrum(1999)]{RaSu99}{{Randall}, L. and {Sundrum}, R.} 1999, Phys. Rev. Lett., 83,
{3370-3373}

\bibitem[Rezzolla et.al.(2001a)]{ram01a}{{Rezzolla}, L., {Ahmedov}, B.~J. and {Miller}, J.~C.} 2001a,
 {\mnras}, 322, {723-740}; Erratum 338, {816-816} (2003)

\bibitem[Rezzolla et.al.(2001b)]{ram01b}{{Rezzolla}, L., {Ahmedov}, B.~J. and {Miller}, J.~C.} 2001b, Found. Phys., 31,
{1051-1063}

\bibitem[Rezzolla \& Ahmedov(2004)]{ra04}{{Rezzolla}, L. and {Ahmedov}, B.~J.} 2004, {\mnras}, 352,
{1161-1179}

\bibitem[Ruffini \& Treves(1973)]{rt73}{{Ruffini}, R. and {Treves}, A.} 1973, Astrophys. Lett., 13,
{109-111}

\bibitem[Schee \& Stuchl\'{i}k(2009)]{sstuch09}{{Schee}, J. and {Stuchl\'{i}k}, Z.} 2009, Gen. Relativ. Gravit., 41,
{1795-1818}

\bibitem[Schee \& Stuchl\'{i}k(2009)]{ssstuch09}{{Schee}, J. and {Stuchl\'{i}k}, Z.} 2009, Int. J. Mod. Phys. D, 18,
{983-1024}

\bibitem[Stuchl\'{i}k \& Kotrlov\'{a}(2009)]{stuch09}{{Stuchl\'{i}k}, Z. and {Kotrlov\'{a}}, A.} 2009,
Gen. Relativ. Gravit., 41, {1305-1343}

\bibitem[Wasserman \& Shapiro(1983)]{ws83}{{Wasserman}, I. and {Shapiro}, S.~L.} 1983, {\apj}, 265,
{1036-1046}

\bibitem[Zanotti \& Rezzolla(2002)]{zr}{{Zanotti}, O. and {Rezzolla}, L.} 2002, {\mnras}, 331,
{376-388}

\end{thebibliography}
\end{document}